\documentclass{article}
\usepackage{amsmath}
\usepackage{cite}
\usepackage{graphicx}
\usepackage{dcolumn}

\begin{document}

\date{}
\title{On some one-dimensional quantum-mechanical models with a delta-potential
interaction}
\author{Francisco M. Fern\'{a}ndez\thanks{%
fernande@quimica.unlp.edu.ar} \\
INIFTA, DQT, Sucursal 4, C. C. 16, \\
1900 La Plata, Argentina}
\maketitle

\begin{abstract}
We discuss a systematic construction of dimensionless quantum-mechanical
equations. The process reduces the number of independent model parameters to
a minimum and, at the same time, provides the natural units of length,
energy, etc. in a clear, straightforward way. We compare this systematic
procedure with the widely adopted one that consists of setting $\hbar=1$. As
illustrative examples, we choose some simple one-dimensional models proposed
recently for the study of localized states in inhomogeneous media.
\end{abstract}

\section{Introduction}

\label{sec:intro}

In a recent paper Savotchenko\cite{S23} discussed ``localized
states in the constant-index and graded-index media separated by
thin defect layer'' in terms of exactly solvable one-dimensional
quantum-mechanical models. In particular he considered the
influence of the intensity of local interaction the excitations
with the interface in short-range approximation on the field
localization'' and obtained ``the exact dependencies of
localization energy
on the interface interaction intensity''. In order to solve the Schr\"{o}%
dinger equation Savotchenko chose some sort of ``conventional
dimensionless units'' based on setting $\hbar =1$. As a result he
was forced to give values to six model parameters in order to
obtain his results. In this paper we resort to a well known recipe
for deriving suitable dimensionless equations\cite{F20} and
compare both kind of calculations.

In section~\ref{sec:gen_mod} we derive a dimensionless equation for a
sufficiently general quantum-mechanical model and obtain the quantization
condition. In section~\ref{sec:first_example} we consider a linear potential
and in section~\ref{sec:other_examples} the parabolic and exponential
potentials. Finally, in section\ref{sec:conclusions} we review the main
results and draw conclusions.

\section{General model}

\label{sec:gen_mod}

The models discussed by Savotchenko\cite{S23} are particular cases
of the one-dimensional Schr\"{o}dinger equation
\begin{equation}
\psi ^{\prime \prime }(x)=\frac{2m}{\hbar ^{2}}\left[ V(x)-E\right] \psi (x).
\label{eq:Schro}
\end{equation}
In order to obtain a dimensionless equation we define $q=x/L$,
where $L$ is the unit of length defined conveniently for each
model\cite{F20}. We thus obtain the eigenvalue equation
\begin{eqnarray}
\varphi ^{\prime \prime }(q) &=&2\left[ v(q)-\epsilon \right] \varphi (q),
\nonumber \\
\varphi (q) &=&\psi (Lq),\;v(q)=\frac{mL^{2}}{\hbar ^{2}}V(Lq),\;\epsilon =%
\frac{mL^{2}}{\hbar ^{2}}E.  \label{eq:Schro_dim}
\end{eqnarray}

We are interested in potential-energy functions of the form
\begin{equation}
V(x)=U_{0}\delta (x)+U(x),  \label{eq:V(x)_delta_gen}
\end{equation}
where $\delta (x)$ is the Dirac delta function. The dimensionless potential
becomes
\begin{eqnarray}
v(q) &=&u_{0}\delta (q)+u(q),  \nonumber \\
u_{0} &=&\frac{mL}{\hbar ^{2}}U_{0},\;u(q)=\frac{mL^{2}}{\hbar ^{2}}U(Lq).
\label{eq:v(q)_delta_gen}
\end{eqnarray}
The delta interaction forces a discontinuity of the first derivative of the
solution at origin
\begin{equation}
\int_{0^{-}}^{0^{+}}\varphi ^{\prime \prime }(q)\,dq=\varphi ^{\prime
}(0^{+})-\varphi ^{\prime }(0^{-})=2u_{0}\varphi (0).  \label{eq:BC_delta}
\end{equation}

In particular we focus on the piecewise potential
\begin{equation}
U(x)=\left\{
\begin{array}{c}
U_{L}(x),\;x<0 \\
U_{R}(x),\;x>0
\end{array}
\right. ,  \label{eq:U(x)_piecewise}
\end{equation}
that leads to
\begin{eqnarray}
\varphi _{L}^{\prime \prime }(q) &=&2\left[ u_{L}(q)-\epsilon \right]
\varphi _{L}(q),\;\lim\limits_{q\rightarrow -\infty }\varphi _{L}(q)=0,
\nonumber \\
\varphi _{R}^{\prime \prime }(q) &=&2\left[ u_{R}(q)-\epsilon \right]
\varphi _{R}(q),\;\lim\limits_{q\rightarrow \infty }\varphi _{R}(q)=0,
\nonumber \\
u_{L}(q) &=&\frac{mL^{2}}{\hbar ^{2}}U_{L}(Lq),\;u_{R}(q)=\frac{mL^{2}}{%
\hbar ^{2}}U_{R}(Lq).  \label{eq:Schro_L_R}
\end{eqnarray}
It follows from $\varphi _{L}(0)=\varphi _{R}(0)=\varphi (0)$ and equation (%
\ref{eq:BC_delta}) that
\begin{equation}
\frac{\varphi _{R}^{\prime }(0)}{\varphi _{R}(0)}-\frac{\varphi _{L}^{\prime
}(0)}{\varphi _{L}(0)}=2u_{0}.  \label{eq:BC_delta_piecewise}
\end{equation}

If we choose $U_{L}^{\prime }=0$ and $U_{R}^{\prime }\neq 0$ we have
\begin{equation}
\frac{\varphi _{L}(q)}{\varphi _{L}(0)}=e^{\alpha q},\;\alpha =\sqrt{2\left(
u_{L}-\epsilon \right) },  \label{eq:Ex1_sol_L}
\end{equation}
and the quantization condition
\begin{equation}
\frac{\varphi _{R}^{\prime }(0)}{\varphi _{R}(0)}-\alpha =2u_{0}.
\label{eq:Ex1_BC_delta}
\end{equation}

\section{First example}

\label{sec:first_example}

One of the simplest potentials that we can choose for $x>0$ is
\begin{equation}
U_{R}(x)=A+Bx,\;B>0,  \label{eq:Ex1_U_R}
\end{equation}
that leads to
\begin{equation}
u_{R}(q)=\frac{mL^{2}}{\hbar ^{2}}A+\frac{mL^{3}}{\hbar ^{2}}Bq.
\end{equation}
Upon choosing $L^{3}=\hbar ^{2}/(2mB)$ and defining $a=mL^{2}A/\hbar ^{2}$
we have $u_{R}(q)=a+q/2$ and the corresponding right eigenvalue equation
\begin{equation}
\varphi _{R}^{\prime \prime }(q)-\left( \delta +q\right) \varphi
_{R}(q)=0,\;\delta =2a-2\epsilon ,  \label{eq:Ex1_eig_eq_R}
\end{equation}
that can be transformed into the Airy equation\cite{AS72}
$Y^{\prime \prime
}(z)-zY(z)=0$ for the variable $z=q+\delta $. Therefore, the solution for $%
q>0$ is
\begin{equation}
\varphi _{R}(q)=Ai(q+\delta ),  \label{eq:Ex1_sol_R}
\end{equation}
and the quantization condition becomes
\begin{equation}
\frac{Ai^{\prime }(\delta )}{Ai(\delta )}=2u_{0}+\sqrt{\delta +\gamma }%
,\;\gamma =2\left( u_{L}-a\right) .  \label{eq:Ex1_quant_cond}
\end{equation}
Note that the solutions depend on only two relevant model
parameters, $u_{0}$ and $\gamma $. This is one of the advantages
of using suitable dimensionless equations\cite{F20}. On the other
hand, Savotchenko\cite{S23} was forced to give values to $m$ and
five model parameters.

\begin{figure}[tbp]
\begin{center}
\includegraphics[width=9cm]{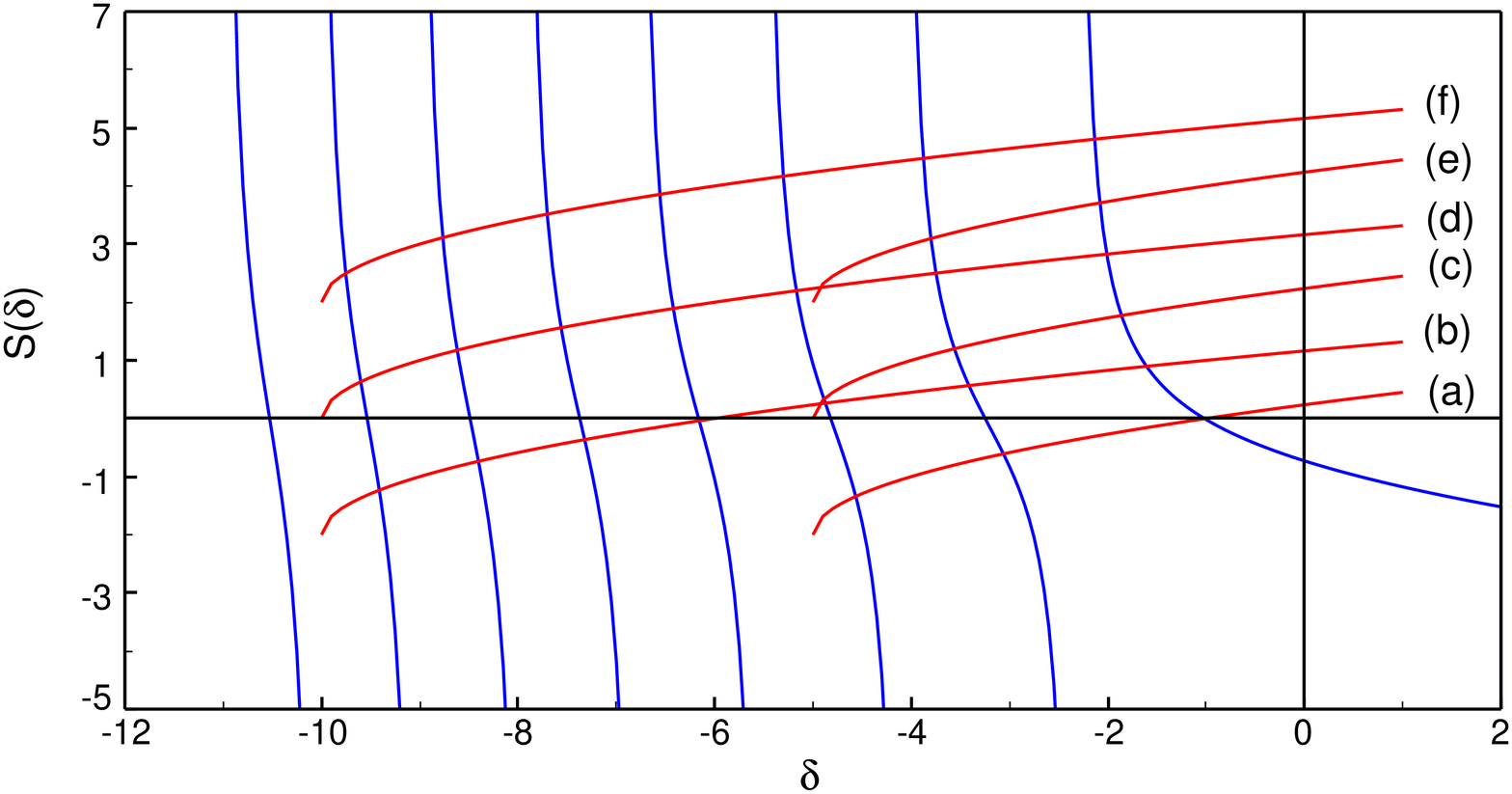}
\end{center}
\caption{$S(\delta)=A_i^{\prime}(\delta)/A_i(\delta)$ (blue line) and $%
S(\delta)=u_0+\protect\sqrt{\delta+\gamma}$ (red line) for (a) ($u_0=-1$, $%
\gamma=5$), (b) ($u_0=-1$, $\gamma=10$), (c) ($u_0=0$, $\gamma=5$), (d) ($%
u_0=0$, $\gamma=10$), (e) ($u_0=1$, $\gamma=5$), (f) ($u_0=1$, $\gamma=10$)}
\label{Fig:Ex1}
\end{figure}

Figure~\ref{Fig:Ex1} shows both sides of equation (\ref{eq:Ex1_quant_cond})
for some values of $u_{0}$ and $\gamma $. The intersections give us the
values of $\delta $ for the bound states. We appreciate that the number of
bound states decreases with $u_{0}$ and increases with $\gamma $. There are
bound states even in the absence of the delta potential (cases (c) and (d))
because $U(x)$ is a well when $\gamma >0$.

For a given suitable pair of values of $u_{0}$ and $\gamma $ we obtain a
series of roots $\delta _{j}$, $j=1,2,\ldots ,k$ from which we derive the
dimensionless energies and eigenfunctions
\begin{eqnarray}
\epsilon _{j} &=&a-\frac{\delta _{j}}{2},  \nonumber \\
\varphi _{j}(q) &=&\varphi _{j}(0)\left\{
\begin{array}{l}
e^{\alpha _{j}q},\;\alpha _{j}=\sqrt{\delta _{j}+\gamma },\;q<0 \\
Ai\left( q+\delta _{j}\right) /Ai\left( \delta _{j}\right) ,\;q>0
\end{array}
\right. .  \label{eq:Ex1_solutions}
\end{eqnarray}
For example, when $u_{0}=1$ and $\gamma =10$ we have: $\delta
_{1}=-2.136182406$, $\delta _{2}=-3.877567571$, $\delta _{3}=-5.301530405$, $%
\delta _{4}=-6.557969140$, $\delta _{5}=-7.703603791$, $\delta
_{6}=-8.766062952$ and $\delta _{7}=-9.753634587$. Note that we
only set two independents parameters, $u_{0}$ and $\gamma $, and
obtain solutions for an infinite set of values of $m$, $U_{L}$,
$A$ and $B$. This is one of the great advantages of using
dimensionless equations\cite{F20}. This results suggest
that the role of the thickness parameter\cite{S23} (which is here absorbed into $%
u_{0}$ and $\gamma $, through $B$) may not be so relevant for the discussion
of the bound states.

\section{Other examples}

\label{sec:other_examples}

For the second example we consider a parabolic potential
\begin{equation}
U_{R}(x)=A+Bx^{2},  \label{eq:Ex2_U_R}
\end{equation}
and choose the unit of length $L=\hbar ^{1/2}/\left( mB\right) ^{1/4}$ so
that
\begin{eqnarray}
u_{R}(q) &=&a+q^{2},  \nonumber \\
a &=&\frac{mL^{2}}{\hbar ^{2}}A=\sqrt{\frac{m}{B}}\frac{A}{\hbar }.
\label{eq:Ex2_u_R}
\end{eqnarray}
The eigenvalue equation for $q>0$ becomes
\begin{equation}
\varphi _{R}^{\prime \prime }(\delta ,q)-\left( \delta +2q^{2}\right)
\varphi _{R}(\delta ,q)=0,\;\delta =2(a-\epsilon ),  \label{eq:Ex2_eig_eq_R}
\end{equation}
that leads to the quantization condition
\begin{equation}
\frac{\varphi _{R}^{\prime }(\delta ,0)}{\varphi _{R}(\delta ,0)}=2u_{0}+%
\sqrt{\gamma +\delta },\;\gamma =2\left( u_{L}-a\right) .
\label{eq:Ex2_quant_cond}
\end{equation}
The solution $\varphi _{R}(\delta ,q)$ can be expressed in terms
of the Whittaker functions\cite{S23,AS72} but the explicit
calculation is not relevant for present purposes. The main point
is that, as in the preceding example, there are only two
independent parameters that completely determine both the
eigenvalues and eigenfunctions. On the other hand,
Savotchenko\cite{S23} was forced to give values to $m$ and five
model parameters.

The third model is given by the exponential potential
\begin{equation}
U_{R}(x)=A+Be^{\beta x},\;B>0,\;\beta >0.  \label{eq:Ex3_U_R}
\end{equation}
In this case we arbitrarily choose $L=1/\beta $ so that
\begin{equation}
u_{R}(q)=a+be^{q},\;a=\frac{mA}{\hbar ^{2}\beta ^{2}},\;b=\frac{mB}{\hbar
^{2}\beta ^{2}},  \label{eq:Ex3_u_R}
\end{equation}
and
\begin{equation}
\varphi _{R}^{\prime \prime }(b,\delta ,q)-\left( \delta +2be^{q}\right)
\varphi _{R}(b,\delta ,q)=0,\;\delta =2(a-\epsilon ).
\label{eq:Ex3_eig_eq_R}
\end{equation}
The solution to this equation can be expressed in terms of Bessel
functions\cite{S23,AS72} and the quantization condition becomes
\begin{equation}
\frac{\varphi _{R}^{\prime }(b,\delta ,0)}{\varphi _{R}(b,\delta ,0)}=2u_{0}+%
\sqrt{\gamma +\delta },\;\gamma =2\left( u_{L}-a\right) .
\label{eq:Ex3_quant_cond}
\end{equation}
In this case we have three independent parameters that completely
determine the eigenvalues and eigenfunctions. On the other hand,
Savotchenko\cite{S23} was forced to give values to $m$ and five
model parameters.

\section{Conclusions}

\label{sec:conclusions}

Throughout this paper we have shown that the appropriate
construction of dimensionless equations simplify the problem
enormously providing the actual units of length, energy, etc. in a
clear, straightforward way . In the examples discussed above the
dimensionless eigenvalue equations exhibit either two or three
independent model parameters instead of the six parameters that
remain when one simply sets $\hbar =1$\cite{S23}. Other
illustrative examples of dimensionless equations in
nonrelativistic quantum mechanics and quantum chemistry are
discussed elsewhere\cite{F20}.

\end{document}